# The Multilayer Nature of Ecological Networks

Shai Pilosof*[1], Mason A. Porter[2,3,4], Mercedes Pascual[1], Sonia Kéfi[5]

[1]Department of Ecology and Evolution, University of Chicago, 1103 E 57 St., Chicago, 60637, USA

[2]Oxford Centre for Industrial and Applied Mathematics, Mathematical Institute, Oxford OX2 6GG, UK

[3]CABDyN Complexity Centre, University of Oxford, Oxford OX1 1HP, UK

[4]Department of Mathematics, University of California, Los Angeles, Los Angeles, California 90095, USA

[5]Institut des Sciences de l'Evolution, Université de Montpellier, CNRS, IRD, EPHE, CC 065, Place Eugène Bataillon, 34095 Montpellier Cedex 05, France

Corresponding author: Pilosof, S. (`pilosofs@uchicago.edu`)

# Preface

Ecological systems are complex and multifaceted in the nature of their interactions. For example, species interact in diverse ways with multiple partners, whose identity can change in space and time. Although networks have provided fundamental representations of ecological complexity, they require extension to systematically and simultaneously capture these multifaceted interactions. We build on recent advances in network science to provide theoretical and practical guidelines on how to tackle multidimensionality in ecological networks with multiple "layers". Multilayer networks hold valuable but largely unrealized potential to complement current representations and provide novel insights into the structure, functioning and dynamics of ecological systems.

# Abstract

Although networks provide a powerful approach to study a large variety of ecological systems, their formulation does not typically account for multiple interaction types, interactions that vary in space and time, and interconnected systems such as networks of networks. The emergent field of 'multilayer networks' provides a natural framework for extending analyses of ecological systems to include such multiple layers of complexity, as it specifically allows one to differentiate and model 'intralayer' and 'interlayer' connectivity. The framework provides a set of concepts and tools that can be adapted and applied to ecology, facilitating research on high-dimensional, heterogeneous systems in nature. Here, we formally define ecological multilayer networks based on a review of previous and related approaches, illustrate their application and potential with analyses of existing data, and discuss limitations, challenges, and future applications. The integration of multilayer network theory into ecology offers largely untapped potential to further address ecological complexity, to ultimately provide new theoretical and empirical insights into the architecture and dynamics of ecological systems.

# Introduction

Networks provide a powerful approach to explore ecological complexity and have generated numerous insights into the understanding of the structure, function, and dynamics of ecological systems[1–7]. Although ecological networks have become fundamental to ecological theory, they have for the most part been studied in 'isolation', as disconnected from other networks, defined at a single point in space and time, and/or aggregated over multiple spatial locations and times. However, natural systems typically exhibit multiple facets of complexity, such as pollinators interacting with flowers in one season but not in another,[8] and the same plant species interacting with both pollinators and herbivores[9,10]. Despite the recognized need to extend the investigation of *monolayer* networks[10,11], there are challenges to doing so explicitly and within a unified framework. Specific examples include the detection of community structure in networks with edges representing different interaction types[10,12] and the analyses of resource flow in temporal networks[13].

Recent advances in the theory of *multilayer networks*[14,15] provide a promising approach. A mathematical framework for the analysis of multilayer networks has been developed only recently, although multilayer network structures, which encode different types of interactions and/or entities as a single mathematical object, have a long history in subjects like sociology and engineering[14,16]. The study of multilayer networks is providing new insights into diverse scientific areas[14,15], including a better understanding of the properties of multimodal transportation networks[17], effects of coupled infection processes on disease dynamics[18], and changes of functional brain networks during the learning of a motor task[19], among others. These advances, in concert with the growing availability of large ecological data sets, provide an exciting opportunity for their theoretical and practical integration into network ecology. In this paper, we define ecological multilayer networks, give examples of the kinds of insights they can enable, and discuss challenges and future applications.

# Ecological Multilayer Networks

Multilayer networks encompass two or more 'layers', which can represent different types of interactions, different communities of species, different points in time, and so on (Fig. 1). Dependency across layers results from ecological processes that affect multiple layers. For example, dispersal of individuals between two patches affects the network structure of both patches[20]. A multilayer network consists of (i) a set of *physical nodes* representing entities (e.g., species); (ii) a set of *layers*, which can include multiple *aspects* of layering (e.g., both time-dependence and multiple types of relationships); (iii) a set of *state nodes*, each of which corresponds to the manifestation of a given physical node on a specific layer; and (iv) a set of (weighted or unweighted) edges to connect the state nodes to each other in a pairwise fashion. The edge set includes both the familiar *intralayer* edges and the *interlayer* ones, which connect state nodes across layers. We provide a formal definition with a detailed example in Box 1.

Previous studies of ecological multilayer networks (Supplementary Table 1) have predominantly used multiple but independent networks of the same system, with interlayer edges formally absent. In such cases, network diagnostics are calculated independently for each layer. For instance, Olesen *et al.*[8] reported—using a plant–pollinator system sampled over 12 years and represented with 12 individual networks—that connectance (i.e., edge density) exhibits little variation over time despite significant turnover of species and interactions. By contrast, networks with explicit interlayer connectivity enable one to address questions about interactions between the processes that operate within and among layers. To set the stage for ecological development, we first identify the major types of layering that are relevant for ecological systems (Fig. 1 and Supplementary Table 1).

## Layers Defined Across Space or Time

Early work on spatial and temporal networks focused primarily on how the composition of species changes in time (e.g., across seasons) or over environmental gradients, especially in food webs[21–26]. More recent studies have focused on studying spatial and temporal dissimilarities in species and interactions[27–29]. This variability finds a natural representation in multilayer networks: one can define a monolayer network at each point in space or time, and then use interlayer edges to connect each node to its counterparts in different layers. Layers in temporal multilayer networks are typically ordered ('ordinal coupling'; Fig. 1a), but the order of the layers is not important for spatial networks ('categorical coupling'; Fig. 1b). Another approach is to use a network of networks. For example, Gilarranz *et al.*[20] defined a spatial network of plant–pollinator networks in which each community of plants and pollinators is a layer, and in which interlayer edges represent species extinction and colonization. By considering a network of layers (see the inset of Fig. 1f), in which each layer is construed as a node, the authors demonstrated an association between the importance of communities (quantified with a measure of node betweenness centrality) and their architecture: communities with higher betweenness in the network of layers are also more nested, with potential consequences for the local stability of the communities[30,31].

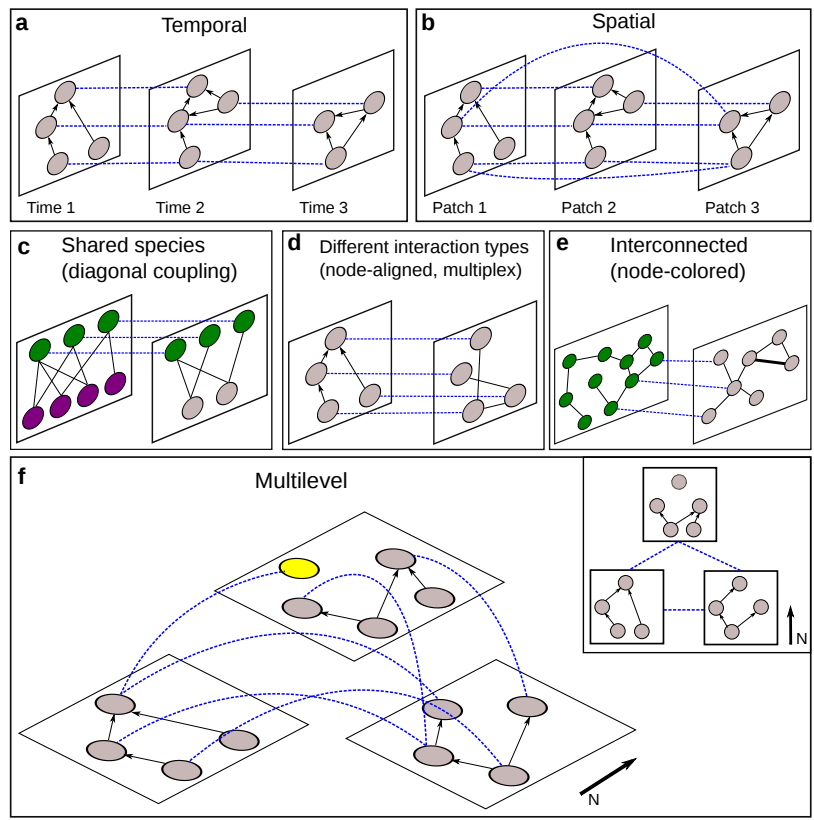

**Figure 1: Multilayer networks in ecology**. In these toy examples, layers are squares, solid black lines are intralayer interactions, and dashed blue arcs are interlayer edges. **a**, Temporal food webs. Nodes are connected to themselves across layers in an ordinal way (one layer follows another). **b**, Spatial food webs. Nodes are connected to themselves across all layers (i.e., categorical interconnections). In panels **a** and **b**, it is permissible for species or intralayer interactions to appear in one layer but not in another. **c**, Networks with different interaction types that are connected through shared species. Layers are 'diagonally-coupled', so interlayer edges occur only between shared species. **d**, One can represent different interaction types among a given set of species (e.g., trophic and facilitative) using a node-aligned multiplex network, in which all nodes appear in all layers and each layer corresponds to a different interaction type. Nodes are connected to all of their counterparts across layers. **e**, Two interacting populations of different hosts. Nodes are individuals, and intralayer and interlayer edges are, respectively, social ties within and between populations. In this example, each node appears in one layer. **f**, A multilevel network representing a metacommunity. Layers are communities, and nodes are species. Intralayer edges are trophic interactions, and interlayer edges represent species dispersal among communities. A species can also disperse to a new community (represented by the yellow node). Communities are often associated with some space (e.g., with different habitats or patches). In the inset, we illustrate that such networks are sometimes represented without explicit specification of interlayer edges. Interactions among species at a lower level automatically impose interactions among the communities.

## Layers Defined by Interaction Type

Because the stability and function of ecological networks can depend on the way in which different interaction types are combined in communities, considering only a single interaction type can give an incomplete picture of system properties[10–12,31–35]. For instance, Bastolla et al.[32] illustrated that both the structure of mutualistic networks and competition for common resources can determine the number of co-existing species in a system. In another example, addition of facilitative interactions to a resource–consumer model affects total system biomass and hence the way an ecosystem functions[11,36]. Finally, Rudolf and Lafferty[37] showed that food webs that include edges representing ontogenetic shifts in addition to trophic interactions change the robustness of the system to extinctions.

Ideally, it is desirable to simultaneously consider the structure of multiple interaction types, and this is achievable with multilayer networks. Some ecological networks have a common set of species (e.g., a set of plants connected to their pollinators and herbivores[9]). One can represent such data[9,38,39] in a multilayer formalism using a 'diagonally coupled' multilayer network (Fig. 1c). Each ecological interaction occurs in a different layer, and interlayer edges connect common species to their counterparts in other layers.

Another approach is to examine different interaction types between all species in a system using 'node-aligned' multilayer networks, in which all entities exist on all layers[12,35,40,41] (Fig. 1d). Kéfi et al.[12,35] used a highly-resolved ecological community from the central intertidal coast of Chile to construct a multilayer network in which each layer includes all species of the community but represents different interaction types: trophic, non-trophic positive (e.g., refuge provisioning), and non-trophic negative (e.g., predator interference). They found that the distribution of non-trophic edges throughout the food web was different from what would be expected by chance (by shuffling the non-trophic edges while fixing the trophic web), suggesting that there is a strong association between the different layers of the network[35]. Such structural patterns suggest the possibility of important dynamic constraints on the combined architecture of trophic and non-trophic interactions[12,35].

## Layers Defined by Different Group Identity

An intuitive way to describe and examine variation in individual-based interactions between populations of the same or different species is with an interconnected network in which each node appears only in one layer (Fig. 1e). In disease ecology, this representation models interpopulation or interspecific disease transmission (when each layer is a population of a different species) at the same time that it considers underlying social networks[42]. For example, intralayer edges can represent the social structure of bat groups, and interlayer edges, transmission of a vector-borne disease among these populations. Moreover, interconnected networks are not limited to individual organisms. For example, one can define a multilayer network in which each layer represents a food web (with its own trophic interactions), and interlayer edges represent trophic interactions among species from different food webs[43,44].

### Layers Defined by Levels of Organization

Biological processes at any given level of organization (e.g., genes, individuals, populations, etc.) can depend on processes at other levels[45]. For example, changes in species' biomass can affect the stability of food webs in dynamical models based on allometry[46]. When the layers in a multilayer network represent different levels of organization, one has a 'multilevel' network, and interactions among nodes at a lower level automatically entail interactions at upper levels[14]. For example, a trophic interaction between two species from two different patches implies that there is an interaction between the patches. The simplest example is a 2-level multilevel network, which can also be construed as a network of networks (Fig. 1f). In an analysis of a 3-level multilevel network (population, community, and metacommunity), Scotti *et al.*[47] illustrated that the metacommunity was sensitive to population-level processes (e.g., social dynamics) that cascaded through different levels. The identification of such dependencies is one of the values of a multilayer approach.

## Analyses of Ecological Multilayer Networks

To illustrate the kinds of insights that can be gained from taking a multilayer approach, we analyse examples of ecological multilayer networks in which layers are explicitly connected. We consider (i) maximum modularity, a structural property that is commonly studied in monolayer networks; and (ii) extinction cascades, a consequence of structure that is common in robustness analyses of networks. We use both synthetic networks and those constructed from empirical data.

**Box 1. Mathematical Definition of a Multilayer Network**

A *multilayer network* is a quadruplet $M = (V_M, E_M, V, L)$. Multilayer networks can have several 'aspects' of layering, and an 'elementary layer' is a single element in one aspect of layering. A 'layer' encompasses one choice of elementary layer for each type of aspect (see the figure for an example). We include such relationships using sequences $\mathbf{L} = \{L_a\}_{a=1}^d$ of sets $L_a$ of elementary layers, where $a$ indexes the $d$ different aspects. Note that $d = 0$ for a monolayer network, $d = 1$ when there is one type of layering, and $d = 2$ when there are two types of layering (as in the figure). The set of entities (i.e., physical nodes) is $V$. The set $V_M \subseteq V \times L_1 \times \cdots \times L_d$ of node-layer tuples (i.e., state nodes) encodes the manifestations of an entity $v \in V$ on a particular layer $l \in \hat{L} = L_1 \times \cdots \times L_d$.

The edge set $E_M \subseteq V_M \times V_M$, which includes both intralayer and interlayer edges, encodes the connections between pairs of state nodes. In a given layer, the intralayer edges encode connections of a specified type (e.g., a certain type of interaction at a given point in time). A function $w: E_M \to \mathbb{R}$ encodes weights on edges. A pair of node-layer tuples, $(u, \alpha)$ and $(v, \beta)$, are 'adjacent' if and only if there is an edge between them. One places a 1 in the associated entry in an adjacency tensor (a generalization of a matrix that consists of a higher-dimensional array of numbers)[14,16] if and only if $((u, \alpha), (v, \beta)) = 1$. Otherwise, one places a 0 in the corresponding entry. One can 'flatten' such an adjacency tensor into a matrix, called a 'supra-adjacency matrix', with intralayer edges on the diagonal blocks and interlayer edges on the off-diagonal blocks (see Supplementary Fig. 1b).

Constraints on the above general definition restrict the structure of a multilayer network[14]. For example, 'diagonal coupling' (see Fig. 1c) is a constraint in which the only permissible type of interlayer edge is one between counterpart entities on different layers. See Kivelä et al.[14] for additional definitions and important types of constraints on $M$ that produce common types of multilayer networks.

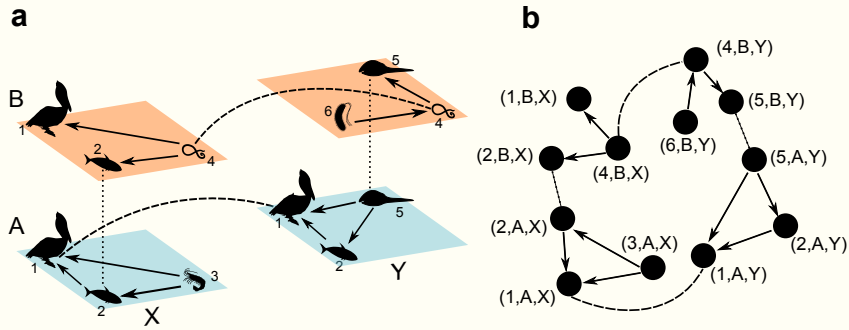

**Toy example of a multilayer network**. **a**, The network has $d = 2$ aspects: (i) different types of ecological interactions, with trophic interactions are in the blue layer ($A$) and host–parasite interactions in the orange layer ($B$); and (ii) space, where $X$ and $Y$ represent different patches. The elementary-layer set for ecological interaction types is $L_1 = \{A, B\}$), and the one for patches is $L_2 = \{X, Y\}$). A layer consists of a tuple of elementary layers. For example, the layer $(A, X)$ encodes trophic interactions at patch $X$. We show intralayer edges using solid arrows. We depict diagonal interlayer edges (e.g., between node 2 on layer $(A, X)$ and node 2 on layer $(B, X)$) with dotted lines; such edges encode the extent to which a parasitized species is more susceptible to predation than a non-parasitized one. Interlayer edges between patches represent dispersal; we show them with dashed arcs. **b**, The 'supra-graph' that corresponds to the multilayer network in panel **a**. Each node in this graph is a node-layer tuple (i.e., a state node) in the corresponding multilayer network. See Supplementary Fig. 1 for an example of how to represent a similar multilayer network as a supra-adjacency matrix. [This figure is modelled after Fig. 2 of Kivelä *et al.*[14].]

## Modularity

In monolayer networks, maximizing modularity can help quantify the extent to which a network is organized into groups (modules) of species that interact more strongly with each other than with other species[48,49]. To our knowledge, modularity has not yet been explored in ecological networks that incorporate interlayer connectivity. To illustrate the distinction between studying a multilayer network and studying a collection of networks, we start with a synthetic example from Fontaine et al.[10]. In this example, a plant–herbivore network and a plant–root-parasite network interconnect via a common set of species, which are the plants (Fig. 2a). Interlayer edges connect each plant species to its counterpart in the other layer (Figs. 1c and 2c,d) and represent the extent to which parasitism affects herbivory. Thus, each plant appears in both layers and has two instances corresponding to different 'state nodes', which can be assigned to different modules. See Supplementary Note 1 for details on how we calculate multilayer modularity and assign state nodes to modules.

The propensity of a state node to belong to distinct modules depends on the relative weights of interlayer and intralayer edges[50,51]. Ecologically, it depends on the extent to which processes in one layer affect those in the other layer. To develop our argument, we consider three conceptual scenarios:

1. When the *interlayer edge weights are* 0, the two networks are indepen-

dent entities. The instances of plants in different layers must belong to different modules, and modules are defined separately for herbivory and parasitism. Hence, herbivory has no effect on parasitism (and vice versa), and no perturbation can pass from one layer to another (Fig. 2b).

2. At the other extreme, *interlayer edge weights are much larger than intralayer ones*. (They are infinite in the limiting case.) In this scenario, herbivory and parasitism always affect each other. This implies that herbivory always renders plants more susceptible to parasites, and vice versa. (Note that it does not imply that each plant is always parasitized and preyed upon.) Consequently, each of the two instances of a plant always belongs to the same module. Note that modules can contain species from any of the three guilds and from either interaction type (Fig. 2d). We also remark that considering infinite interlayer edge weights produces in general different results from aggregating a multilayer network into a monolayer network.

3. *Interlayer and intralayer edge weights have comparable values* (i.e., they are on similar scales). In this scenario, herbivory has some effect on the propensity of a plant to be parasitized, and vice versa. A plant can therefore interact strongly with a given set of herbivores in one layer and with a given set of parasites in another. Each of the two instances of a plant can belong to different modules. Modules can contain species from any of the three guilds and either interaction type, but the identity of the modules can be rather different from those in the previous two cases (Fig. 2c).

This example illustrates that defining community organization in multilayer networks depends strongly on the extent to which the ecological processes that operate in the different layers affect each other (in our example, the relationship between parasitism and herbivory). Another insight is that considering intermediate values of interlayer edge weights provides a possible means to identify the plants that can buffer perturbations: the two instances of such plants would be assigned to different modules (Fig. 2c).

One challenge is to quantify the interlayer edges. In this example, one way to measure the extent to which herbivory and parasitism affect each other is to conduct a series of experiments in which one group of plants of a given species is exposed to herbivores while a second group of the same species (control) is not. For example, if a plant exposed to a given herbivore species is infected by twice the number of parasites compared to the control plant, then the value of the interlayer edge is 2. One can then average across a series of experiments of different herbivore–parasite combinations to obtain the values of the interlayer edges for that plant species.

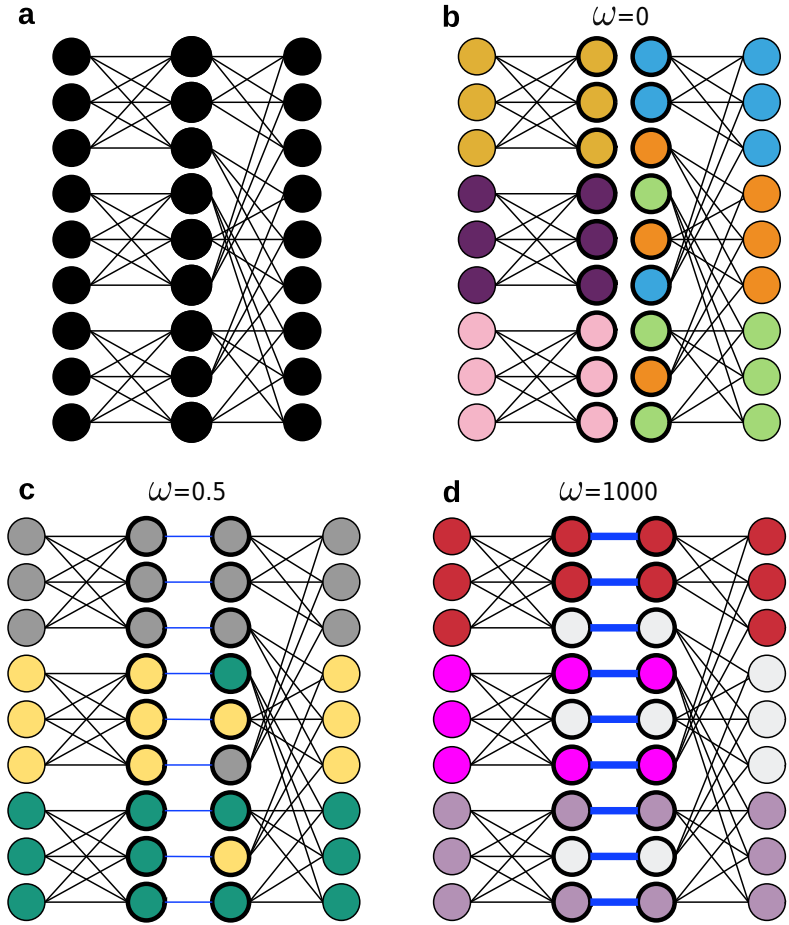

**Figure 2: Modularity maximization in a diagonally-coupled multilayer network.** **a**, An example network (from[10]) that does not have interlayer edges. In panels **b**, **c**, and **d**, the network has two layers, which interconnect via a common set of nodes (with thick borders). Interlayer edges (in blue) connect the two instances of a node. In panels **b**, **c**, and **d**, we test three different scenarios. In each panel, nodes of the same colour belong to the same module. We construe the left set of nodes as the root-parasite guild, the middle set with thick borders as plants, and the right set as herbivores. For our calculations, we use $\omega = 1000$ to approximate interlayer edge weights of $\infty$. In panel **b**, we obtain a maximum modularity of $\overline{Q}_B \approx 0.667$, each layer has 3 modules, and no plant state nodes appear in the same module. In panel **c**, $\overline{Q}_B \approx 0.651$, and the network is partitioned into a mean of $\overline{n} \approx 3.04$ modules. On average, 4.6 (i.e., approximately 51%) of the plants are assigned to more than one module. In panel **d**, $\overline{Q}_B \approx 0.99$, and we obtain a mean of $\overline{n} \approx 3.38$ modules, and state nodes of each plant belong exclusively to a single module. See Supplementary Notes 1 and 2 for the details of our calculations.

Several studies outside of ecology have illustrated that modular (and other mesoscale) network structures can change over time[19,50,52]. In ecology, such network variation was considered in a study based on the analysis of multiple disconnected networks[53]. Time-dependent modular structure, including changes in module composition over time, can also be examined by studying a multilayer network. As an example, consider a network representing the infection of 22 small mammalian host species by 56 ectoparasite species during six consecutive summers in Siberia (1982–1987)[54,55], yielding a multilayer network with 6 layers. We quantify intralayer edge weights as the prevalence of a given parasite on a given host. Interlayer edges connect instances of the same species across consecutive time points (Fig. 1a), representing the relative changes in abundance between two consecutive summers. For example, if a host has an abundance of 10 in one year and 5 in the next, then the value of the interlayer edge is $5/10 = 0.5$ (Supplementary Note 3). The idea behind this choice of interlayer edge values is that temporal fluctuations in abundance affect the availability of hosts to parasites and parasite pressure on hosts, and these factors in turn affect host–parasite interactions in any given time point (i.e., the intralayer edges; see Eq. (3) in Supplementary Note 3). This choice results in (nonuniform) interlayer edge weights that are on a similar scale than that of intralayer edges.

In a temporal network, a given species can interact strongly with some species at one time and with other species at other times. Consequently, each state node can belong to a different module at different times, and modules can vary in size over time[50,51] (Fig. 3). We define 'host adjustability' and 'parasite adjustability' as the proportion of hosts and parasites, respectively, that change module affiliation at least once. We observe nonnegligible values of this measure: about 47% of the hosts and about 35% of the parasites change their module affiliation at least once, and a module's size also changes over time (Supplementary Figs. 2,3). One interpretation of this pattern is that the same species is functionally different at different times. For example, one ecological characteristic of a host is the extent to which it supports populations of different parasite species in a community. Fluctuations in the abundance of different host species (encoded in the interlayer edges) can lead to fluctuations in the availability of hosts to parasites. Additionally, there are temporal changes in interaction patterns among other species in the network (encoded in the intralayer edges). These mechanisms lead to a time-dependent distribution of parasites in hosts in which the same host species supports populations of different parasites at different times. This variation is expressed as the assignment of the same host species to different modules in different layers.

Constructing and analysing a temporal network also allows one to consider hypotheses on the effect of intralayer and/or interlayer connectivity on community structure. This approach dominates studies of monolayer ecological networks in which the structure of an observed network is com-pared to that of networks within an ensemble that have been shuffled in particular ways[4,56] (or to those generated with a generative random-graph model[48,49]). The added value of the generalization of this approach to multilayer networks is that it allows one to test hypotheses that relate directly to the temporal structure of a community. For example, we can shuffle the interlayer edges between each pair of consecutive layers (separately for hosts and parasites) to test the hypothesis that the modular structure is a result of random temporal changes in species abundance (Supplementary Note 3). The observed network has higher maximum modularity than the shuffled networks ($\overline{Q}_B^{\text{observed}} \approx 0.55$ versus $\overline{Q}_B^{\text{shuffled}} \approx 0.21$; the p-value is $P < 0.001$). Additionally, it has about 6 (15) times fewer modules than in networks where the order of interlayer host edges (parasite edges) have been shuffled. All hosts and parasites switch modules at least once in the shuffled networks. These observations lead us to reject the above hypothesis. One can also hypothesize that the modular structure of the community is: (i) a result of random associations between hosts and parasites in any given layer, and (ii) independent of the temporal order in which hosts and parasites are observed (Supplementary Note 3). We reject these hypotheses as well (Supplementary Tables 4,5). The rejection of the three hypotheses improves understanding of the functional groupings of species, by demonstrating that host–parasite interactions are structured nonrandomly in time, depending both on how species interact within a given time period and on their persistence in time as measured by changes in species abundance. Consequently, altering host–parasite interactions or the survival probability of species (e.g., by applying parasite control programs) would strongly affect temporal community organization.

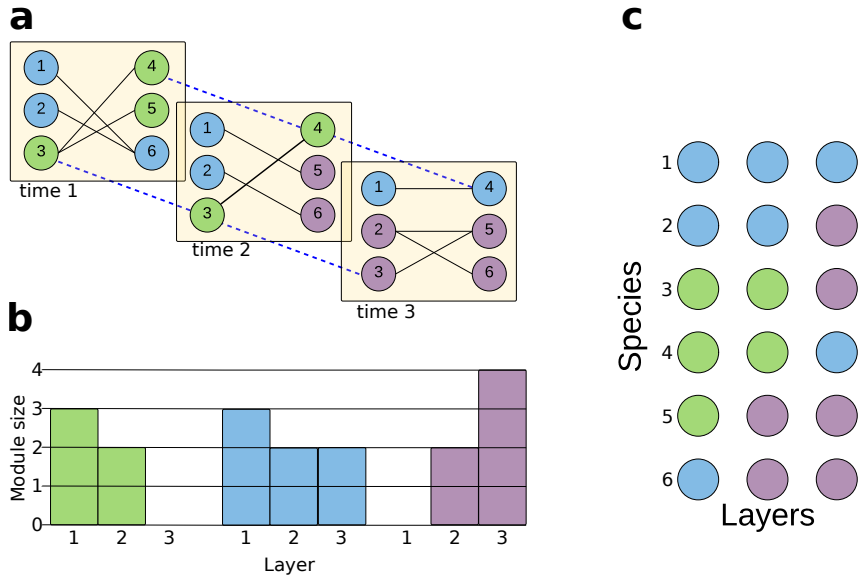

Figure 3: **A toy example of temporal modularity maximization.** The figure illustrates that species can change their module affiliation across time points. We represent six species with numbers and indicate three different modules using different colours. **a**, The toy temporal network has 3 layers, and each layer has a bipartite structure. All species occur in all layers. For clarity, we represent interlayer edges using blue dashed lines for two species. **b**, Modules can change in size (i.e., the number of species that they include ) across layers. For example, the green module does not exist at time point 3, and the purple module does not exist at time point 1. **c**, Representation of the module affiliation of each species in different layers. For example, species 1 does not switch modules, whereas the other species switch once each.

Any assessment of such questions by studying each layer separately would be necessarily incomplete. Modules in independently-examined layers of a network are disconnected and independent from each other. Hence, one cannot directly address the effect of interlayer phenomena (e.g., temporal changes in species abundance or in local network architecture) on modular structure (Supplementary Notes 1, 3). In the host–parasite network without interlayer edges, species assignment to modules in any given layer contains, on average, only about 35% of the information on species assignment to modules in the interconnected temporal network (Supplementary Note 3.3).

As an alternative to using a multilayer network, ecologists can aggregate species and interactions (e.g., across space or time). Aggregation may be necessary when species interactions (intralayer edges) are sampled sparsely in time. However, data aggregation entails a set of (usually implicit) assumptions[14,57], and different aggregation methods can lead to qualitatively different conclusions. For instance, by calculating 'reducibility'[58] (Supplementary Table 2) in the host–parasite network, we can quantify whether some layers contain overlapping information, and can therefore be aggregated. We find that all six layers are necessary to describe the entire complexity of the system (Supplementary Note 3.4). Consistent with this finding, the affiliation of species to modules in the aggregated network provides only about 52% of the information on their affiliation to modules in the multilayer network (Supplementary Note 3.3).

Taken together, our computations illustrate that multilayer networks lend themselves naturally to asking questions about time-dependent phenomena[13,19,50,51]. Temporal variations in the size and composition of modules may be relevant to phenomena such as species coevolution, coexistence, and community stability[4,10,59]. For instance, hosts assigned to more than one module may provide important bridges for transmission of ectoparasites across years and/or groups of strongly connected hosts. These hosts can change their ecological function in a system[55], and such flexibility in community structure may contribute to system robustness in the face of perturbations.

## Robustness to Perturbations

Network structure can affect the stability of ecological communities[7,30,31]. To illustrate how multilayer networks can contribute to studies of stability, we use a network with two layers, plant–flower-visitors and plant–leaf-miner parasitoids[38], that are interconnected via the same set of plants (Fig. 1c, Supplementary Note 4). We investigate the patterns of parasitoid extinctions in two scenarios: (i) direct secondary extinctions due to plant removal; and (ii) tertiary extinctions due to the removal of pollinators, which causes plant secondary extinctions that, in turn, result in parasitoid tertiary extinctions (Supplementary Table 6). A plant (respectively, a parasitoid) goes extinct when it becomes completely disconnected from flower visitors (respectively, flowers). We find that parasitoid extinctions occur more slowly in the multilayer network than in the plant–parasitoid monolayer network. Additionally, allowing plant extinctions in addition to flower visitors extinctions in scenario (ii) leads to nontrivial extinction patterns (Fig. 4). Therefore, considering the multilayer nature of the network changes the qualitative conclusions about the robustness of parasitoids to extinctions.

There are many ways to extend this kind of analysis (e.g., Supplementary Note 4). For example, we considered above the values of the interlayer edges to be infinity, to enforce perfect correspondence between the plant states in the two layers. By relaxing this assumption, one can model for instance, changes in plants abundance as a function of changes in flower visitors or leaf-miners abundance. This can be done, for example, using a dynamical system (described by a set of ordinary differential equations, as has been done recently in the study of a meta-foodweb model[60]) on each layer, and by using the values of the interlayer edges to couple these processes. Analysis of such scenarios should be valuable for understanding the interplay among different interaction types and their effect on system robustness.

## Limitations and Challenges

Whether a multilayer approach is more appropriate than a monolayer one obviously depends on the specific research question. Collecting the necessary data for questions requiring multilayer networks can be resource-intensive, as the data needs to be gathered from multiple places, at multiple times, and/or with different observational methods to capture different types of interactions. Measuring interlayer edge weights may require additional sampling efforts that are different from those used for collecting data on intralayer edges. Fortunately, data sets are already becoming available (Supplementary Table 1), but they are scattered in the literature and would need to be curated.

One challenge is to define the meaning, and measure the values, of interlayer edges, and the choice of definition can itself play a significant role in the analyses. For example, interlayer edges that connect species in two different communities may relate to species dispersal or changes in a species' state (e.g., abundance). Furthermore, intralayer and interlayer edges can represent ecological processes at different scales, and it is not always clear how to define the relative weight of interlayer edges with respect to intralayer edges. How to choose appropriate values for interlayer edges remains a topic of active research in the study of multilayer networks. For some applications (e.g., transportation), there already exist principled ways to choose values, and for others (e.g. social networks) progress is underway. In ecology, this issue remains completely uncharted territory. Where possible, it would be best to measure interlayer edges directly[61], and we illustrated one possibility in our analysis of a temporal network. An *ad hoc* approach, which has been very insightful for several applications outside of ecology[19,51,52,62], is to systematically sweep through a set of values, and assess the robustness of qualitative conclusions in different regions of parameter space, as well as how they change as a function of interlayer edge weights. For example, one can ask how the relative relationship between two interaction types affects the dynamics of an ecological system. For this theoretical question, one would measure a quantity of interest (e.g., some network diagnostic) across different interlayer edge weights. We also note that different types of interactions can also involve different 'currencies'. For example, pollination is measured differently than dispersal, and it is important to consider discrepancies in the scales of the two edge types.

As with previous advances in network ecology, one can borrow and adapt techniques from the more general science of networks[2,4,59,63,64]. Although a natural first step is to adopt (and adapt) diagnostics and algorithms from the many that are available in applications of multilayer networks to other disciplines[14,15] (Supplementary Table 2), it is important to determine which existing methods are best suited for ecological applications, and to develop methods and diagnostics with ecology specifically in mind. These methods include multilayer versions of common network diagnostics (e.g., vulnerability and generality), null models to test hypotheses about network structure[62,65], and mathematical models for dynamical processes. Availability of software[59,66,67] has played an important role in the accessibility of monolayer network analyses, and the same should become the norm for the analysis and visualization of multilayer networks with the development of well-described and easy-to-use tools. Several software packages are already available (Supplementary Table 3), even though these are not yet as mature as those for monolayer ecological networks.

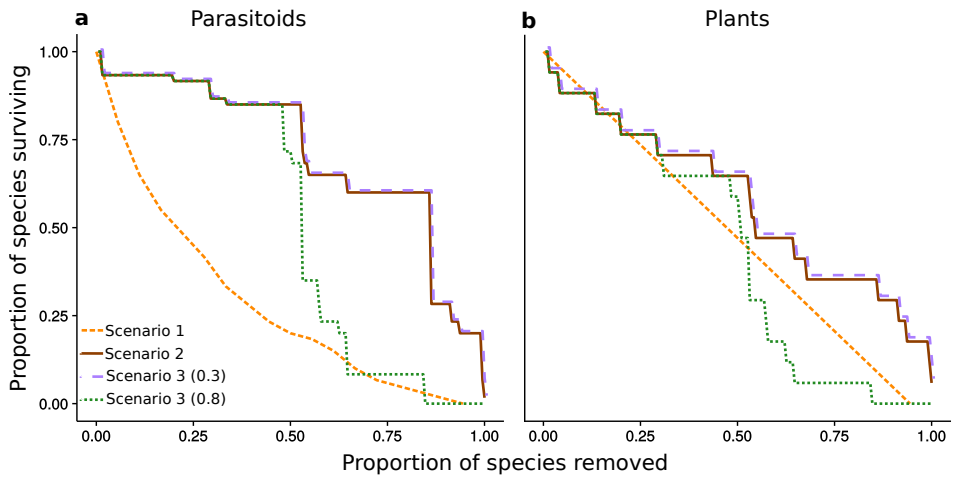

Figure 4: **Network robustness to species removal in a multilayer network of plant–flower-visitors and plant–leaf-miner parasitoids.** The proportion of surviving (**a**) leaf-miner parasitoids and (**b**) plants as a function of the proportion of species removed in three scenarios. Scenario (i) simulates a monolayer case, where we remove plants from the plant–leaf-miner parasitoid layer (orange curves). This results in a linear decrease in the proportion of plants surviving (in panel b). In scenario (2), we remove flower visitors and quantify secondary extinctions (the proportion of surviving species) in plants and concomitant tertiary extinctions of leaf-miner parasitoids (brown curves). Scenario (3) is the same as (2), but with additional plant removal (which is independent of the secondary plant extinctions) with probability 0.3 (purple curves) or 0.8 (green curves). We detail the algorithms for each of the removal scenarios in Supplementary Table 6. The is overlap between the curves for scenarios 2 and 3 with low probability, so we separated them slightly for clarity. In panel **a**, we note a 'transition' in the proportion of species that survive in scenario 3. Although discerning whether this is a general feature of extinction processes in ecological multilayer networks requires further research, a plausible explanation is that when there is a high probability of random independent extinctions of plants, the system reaches a threshold in which too many plants have become extinct and thus can no longer sustain the community.

## Future Directions and Applications

We have argued that multilayer networks provide a versatile and powerful framework for investigations in community ecology. This is also true for biogeography, where networks that include species dispersal or colonization/extinction dynamics as interlayer edges can help one understand how spatial dynamics affect community structure and stability[20]. A multilayer formalism can address not only correlations between geographic distance and beta diversity of species and interactions (which to date have been studied using multiple disconnected networks; see, e.g., Trøjelsgaard et al.[29]), but also the mechanisms behind such correlations. Another open area concerns the effect of space on the stability of ecological communities[68]. To this end, one could use a multilayer network that explicitly incorporates different instances of the same species at different locations instead of a matrix of interactions for a single community in space[30,34]. Habitat alteration is a growing concern, and it would be interesting to explore the stability of ecological networks with different spatial connectivity patterns (encoded in the interlayer edges).

Ecological multilayer networks is a new framework, and the development of theoretical models is critical as they provide a basis for comparison with data[11]. Indeed, a multilayer ecological framework should stimulate further theoretical developments. For instance, it provides a way to consider the structure underlying multiple interaction types in dynamical models, which thus far have been modelled on top of a single web describing a single interaction type[32,34]. Interlayer edges in such models encode the amount of coupling between different dynamical processes and provide a way to encode the relative importances of different interaction types in determining processes such as species' population dynamics. For example, a key question in disease ecology is assessing the roles that different types of host–parasite interactions play in disease transmission (e.g., the contact patterns among individuals[69] or the trophic web within which parasites and hosts are em-

bedded[70]). Using a multilayer network with three layers—contact network, trophic web, and host–vector interactions—Stella et al.[41] modelled the relative importances of vectors for diseases that can be transmitted both via trophic interactions and vectors by varying the coupling between the trophic and the vector layers. They reported that transmission that spreads only on the trophic layer hinders the infection of host populations.

Multilayer networks can also advance metacommunity theory, where interlayer edges provide a way to develop spatially-explicit models to investigate how species move among local communities, thereby creating spatial structure in the regional species' pool. For example, one can study spatial structure in resource and species flow in interconnected food webs, such as those of lakes or ponds that are interconnected via common sets of animals. In such examples, interlayer edge values can represent dispersal[20] or biomass flows among patches. One can also use a temporal food web to explore bioenergetic flows across food webs—an area that remains largely unexplored[13]. Layers can represent temporal instances of a given food web, and interlayer edges can represent changes in species biomass with time.

Another key question in food-web theory is the effect of parasitism on food-web structure and stability[70–72]. For example, parasitism may have different effects than trophic interactions on a given species. Parasitized hosts can be more susceptible to predation, but it is unclear how the structure of host–parasite networks affects the trophic interactions between hosts. One way to model such systems is by coupling a host–parasite network and a food web (Box 1 and Supplementary Fig. 1). Multilayer networks also provide a possible approach for analysing disease transmission when there are multiple hosts and parasites[18,42], which to date has been difficult to study.

In animal behaviour, one can explore networks in which intralayer edges between individuals represent reproduction and interlayer edges represent movement (and hence gene flow), to study genetic relatedness among individuals as a function of both dispersal and within-group social behaviour. In movement ecology, multilayer networks can help model relationships between a network of social interactions and a network of movement patterns to understand moving decisions. Understanding movement has consequences for conservation biology, where one can represent different connectivity scenarios using a multilayer network in which each layer describes the movement patterns of a different species. Such models would inform decision-makers on which land-use designs are best-suited for the movement of a diverse set of species. Multilayer networks can also help improve the identification of keystone species. A simple notion of a keystone species arises from calculating a "centrality" measure of species in a food web (i.e., keystone species have the highest values of that centrality)[73], but a more nuanced definition of centrality—and hence of keystone species—can incorporate participation in several interaction types[61], as well as temporal and spatial dependencies.

Finally, multilayer networks can be used to study reciprocal effects be-tween ecological and non-ecological systems. For example, Baggio et al.[74] represented three indigenous Alaskan communities using a multiplex network in which each layer is a unique combination of ecological resources and social relations. They found that changes to the social relations have greater impact on the robustness of the networks (and hence the human communities) than depletion of ecological resources (e.g., removal of marine species used as food).

## Conclusions

The simultaneous expansion in the availability of ecological data and the tools to analyse multilayer networks[14,15,18,75], provides a timely and valuable opportunity for ecologists to explore the multilayer nature of ecological networks. The strength of a multilayer approach lies in its ability to formulate and analyse complex systems in a way that explicitly incorporates processes operating both within and across layers (as well as the interaction between these processes). Formulating systems as multilayer networks also allows one to address questions that are off-limit based on monolayer representations. Working within the same framework facilitates comparisons of results across ecological systems and network types, because of consistency in technical terms and methodology. A unified framework should further encourage collaboration with scientists from other disciplines. Other fields of study, in turn, would benefit from methodology and theory developed for ecological multilayer networks, as has been the case for monolayer ecological networks[76]. In closing, the integration of multilayer network theory into ecology offers an additional perspective, with the potential to provide new theoretical and empirical insights, into the architecture and dynamics of ecological systems.

## Data and Code

We provide the raw data for the example temporal network in the Supplementary Information (Data set 1). The code for general procedures to prepare data, manipulate networks, post-process modularity-maximization calculations, and analyse network robustness is written in R, and it is available at 10.6084/m9.figshare.3472664. The code for examination of modular structure is written in MATLAB, and it is available at 10.6084/m9.figshare.3472679.

## Acknowledgements

SP was supported by a James S. McDonnell foundation postdoctoral fellowship for the study of complex systems and by a Fulbright postdoctoral fellowship from the US Department of State. MAP was supported by FET-Proactive project PLEXMATH (FP7-ICT-2011-8; grant #317614) funded by the European Commission. We thank Lucas Jeub for help with the analysis of multilayer modularity, Manlio De Domenico for insightful discussions on multilayer networks, and Boris R. Krasnov for help with the temporal network data. MAP thanks Mikko Kivelä and other collaborators for helping to shape his view of multilayer networks and the participants at the first GCEE workshop on Resilience and Recovery of Biological Networks to Environmental Fluctuations for illuminating discussions.

## Author contributions

SP conceived the idea, performed numerical simulations, and analysed the data; MAP contributed insights about the mathematics of multilayer networks and data analysis; and MP and SK contributed insights on the use of multilayer networks in ecology and the interpretation of results. All authors wrote the manuscript.

## Competing financial interests

The authors declare no competing financial interests.

# Supplementary Information

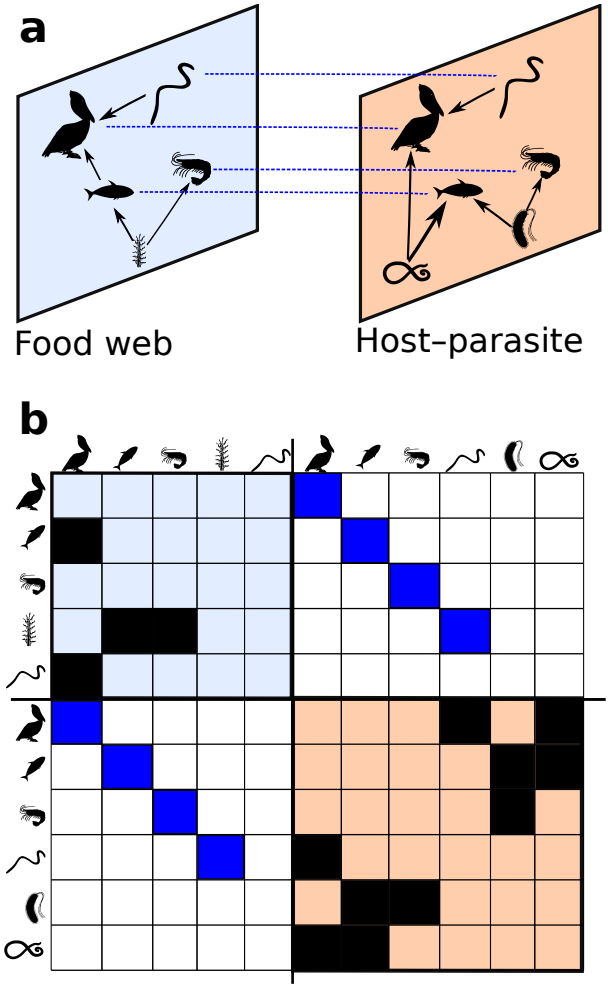

**Supplementary Figure 1: An ecological multilayer network with two interaction types**. **a**, The network depicts (light blue layer) an aquatic food web and (light orange layer) a host–parasite network. Species that occur in both layers are interconnected with dashed blue interlayer edges. Note that one parasite appears in both layers and can therefore play a role as a prey and/or a parasite for the pelican. **b**, A supra-adjacency representation of the network in panel **a**. Coloured blocks correspond to the network layers. We show intralayer and interlayer edges, respectively, with black and blue matrix cells. Species that appear in two layers also appear twice in the matrix. These are different state nodes of the same physical node.

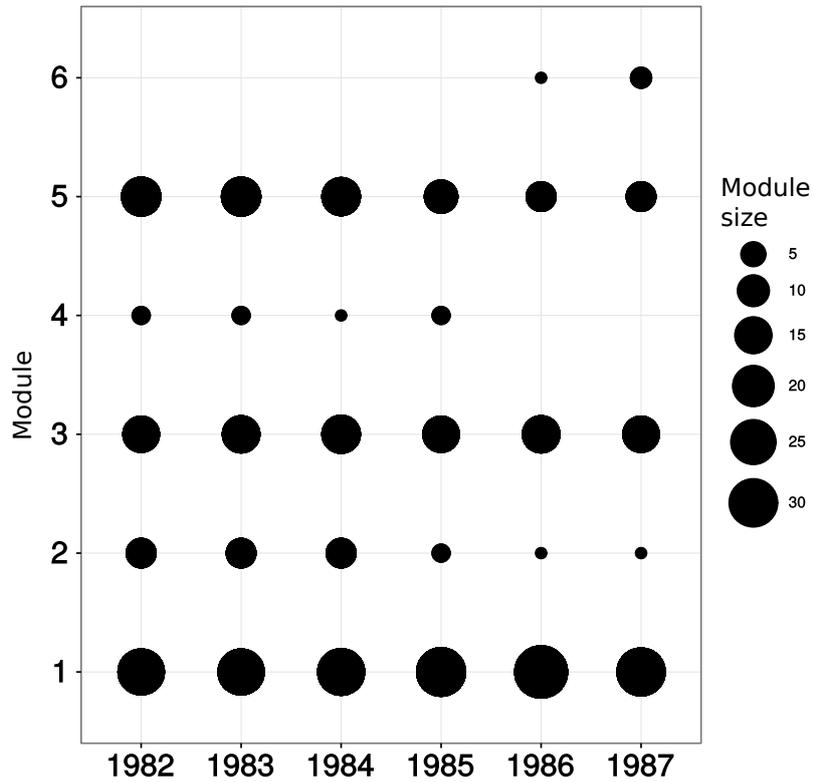

Supplementary Figure 2: **Yearly variation in module size**. We examine modular structure in a network that represents the infection of small mammalian hosts by fleas and mites. Because species can switch modules across years (i.e., across layers), we observe variation in module size. For example, module 1 maintains a relatively large size across years, whereas module 4 exists only in four of the six layers. This figure depicts a particular example among 100 instantiations of the generalized Louvain algorithm[77] for multilayer modularity maximization of the host–parasite network). This instantiation is the one with the maximum value of the maximum modularity. See Supplementary Note 1 for details about our calculations.

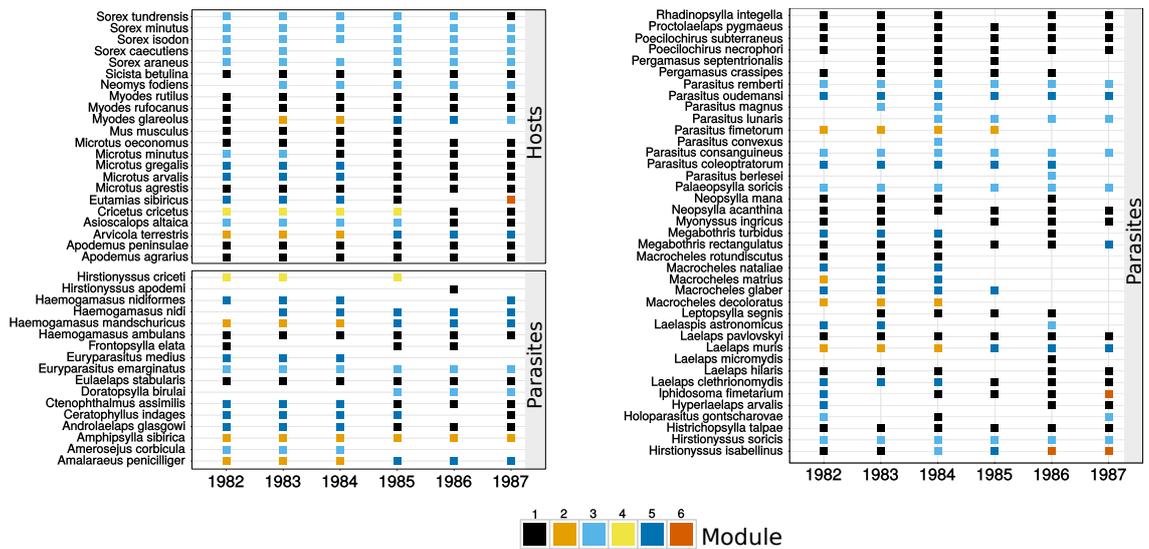

Supplementary Figure 3: **Module affiliation of species across years for the host–parasite network examined in the main text.** We represent different modules using different colours. We show community assignments for a particular example (the one with the highest value of maximum modularity $Q_B$) among 100 instantiations of applying the generalized Louvain algorithm[77] for multilayer modularity maximization. The figure illustrates that species can change their module affiliation across years. Missing data points indicate that a species is not present in a particular year.

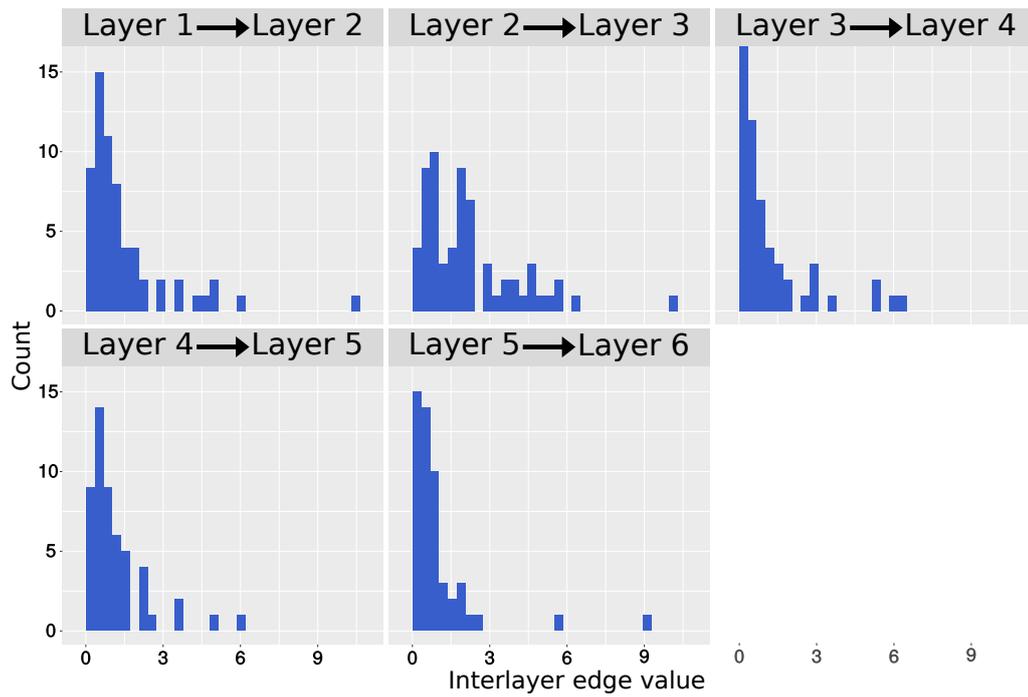

**Supplementary Figure 4: Distribution of the values of interlayer edges of a host–parasite temporal network**. We calculate interlayer edge weights as the relative changes in abundance of a species during the time window $t \to t+1$. Each histogram gives the distribution of values for a given time window of the network (there are 6 layers and hence 5 time windows). See the main text and Supplementary Note 3 for details on the calculation of interlayer edge weights.

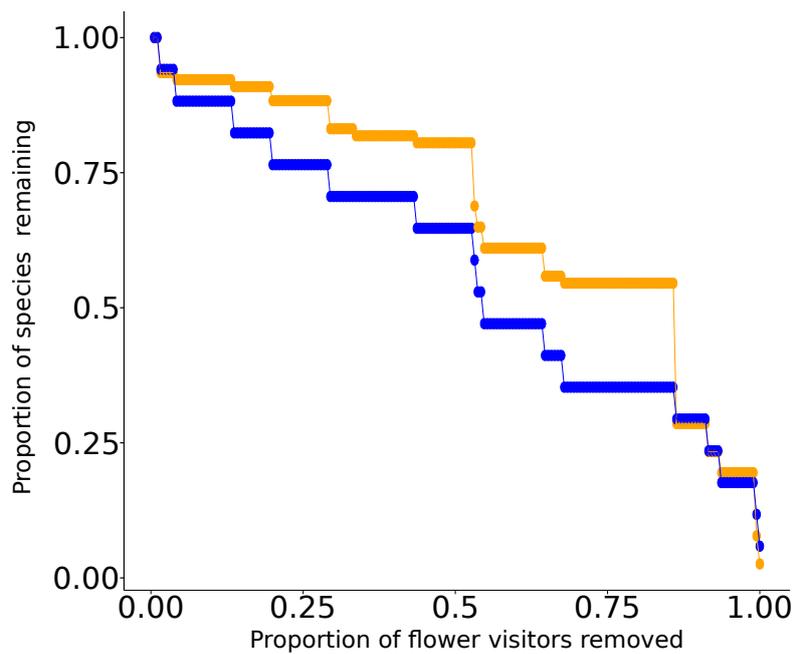

**Supplementary Figure 5: Network robustness to species removal for the plant–flower-visitors and plant–leaf-miner parasitoids diagonally-coupled multilayer network.** The curves indicate the proportion of (blue curve) surviving plants and (orange curve) surviving plants and parasitoids as a function of the proportion of flower visitors that have been removed. In each scenario, we remove flower visitors in the same order, which is by increasing intralayer node degree (that is, we first remove the lowest-degree flower visitors, and we proceed accordingly). See Supplementary Note 4 for more details.

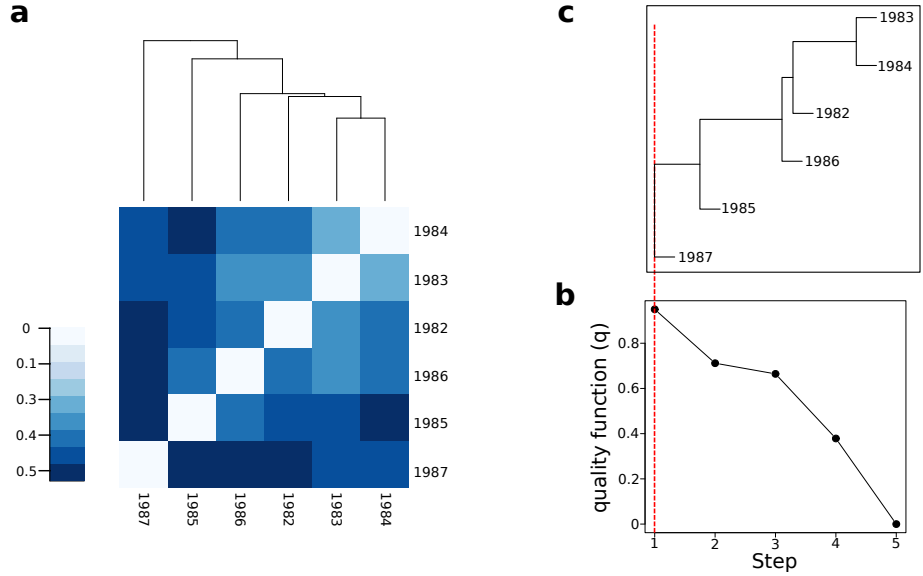

**Supplementary Figure 6: Reducibility of a host–parasite temporal network. a**, Following De Domenico et al.[58], we compute the Jensen–Shannon distance $D_{js}$ between each pair of layers as a proxy for layer redundancy. **b**, We perform hierarchical clustering on the resulting distance matrix. The output is a dendrogram whose leaves represent the initial layers and whose internal nodes indicate layer merging. **c**, At each step, we aggregate the clustered layers corresponding to the smallest value of $D_{js}$, and we quantify the quality of the resulting network in terms of distinguishibility from the aggregated graph (i.e., a graph which includes all physical nodes and in which edges are the sum of intralayer edges across all layers) by a global quality function $q$. (See De Domenico et al.[58] for details on how to calculate $q$.) An 'optimal' partition (which need not be unique) is one for which $q$ is maximal. In this example, we obtain maximal $q$ from the partition (indicated by the dashed red line) in which the 6 layers are separate.

**Supplementary Table 1:** Types of ecological multilayer networks that have been examined and some questions that can be addressed by using them. In our list, we include studies that used multiple networks that were not connected explicitly to each other via interlayer edges.

| Layer | Nodes | Interlayer edges | Intralayer edges | Structural characteristics | Primary research themes/questions | Reference | Figure |
|---|---|---|---|---|---|---|---|
| Time | S | Each node to itself (ordinal) | M, P, T | A set of networks defined separately at different time points. | Temporal variation in network or node diagnostics (e.g., centrality, connectance). Network assembly dynamics. Temporal variation in beta diversity of species and/or interactions. | 8,13,21–23,27,55,78–97 | Fig. 1a |
| Space | S | Each node to itself (categorical) | M | A set of networks defined separately at different sites. | Spatial variation in network or node diagnostics (e.g., number of species, connectance). Spatial variation in beta diversity of species and their interactions as a function of geographic distance. | 21,22,27,29,79,98–100 | Fig. 1b |
| Interaction type | S, FA | Each node to itself | M, P, T, G | Networks interconnected via shared species (diagonal coupling). | How do different interaction types change understanding of ecological and evolutionary dynamics? How do different interaction types affect the stability and robustness of a system? What is the association between host immunogenetics and host–parasite interactions? | 9,10,38,39,68,101,102 | Fig. 1c |
| Interaction type | S | Each node to itself | T, P, NT+, NT− | Layers with the same set of species and different interaction types (e.g., parasitism, trophic, facilitation, mutualism) or signed networks. | Quantify architecture of trophic and non-trophic interactions, and quantify the association between these different interaction types. How does including different interaction types affect functional diversity and species roles? How are parasite transmission dynamics affected by different routes of transmission? How do invasive or migratory species affect the communities into which they arrive? | 12,37,40,61,103 | Fig. 1d |
| Community | S | Each node to itself | T | Food webs interconnected via common species. | How does interaction between food webs (via common species) affect the dynamics of each food web separately? | 26,43,104 | Fig. 1c |
| Community | S | Dispersal | M, T | Networks of communities (e.g., food webs, mutualistic bipartite networks) connected by species dispersal. The communities can be spatially embedded. | How are dispersal and colonization-extinction dynamics associated with network architecture in each community? | 20,105–107 | Fig. 1f |
| Multiple | I, P, C | Each node to a node in an adjacent layer. | Sc,T, D | Network layers describe different hierarchical levels of a system. | How do processes at one level affect those in other levels? | 47 | Fig. 1f |
| Population | I | Each node to itself | Sc | Network layers defined separately for each behavioural interaction. | How are the function and dynamics of animal social networks defined by different behaviours? | 108 | Fig. 1d |
| Population | I | Between nodes from different layers | Sc | Layers interconnected through one or more nodes. | Disease dynamics in interconnected populations of different hosts. | 42 | Fig. 1e |

Nodes: Species (S), Individuals (I), Communities (C), Metacommunities (MC), Functional alleles (FA). Intralayer edges: Mutualistic (M), Parasitism (P), Trophic (T), Non-trophic positive (NT+), Non-trophic negative (NT−), Genetic (G), Dispersal (D), Social (Sc). References are to studies that explored either networks with multiple but independent layers or multilayer networks with interlayer edges. The 'Interlayer' column is relevant for studies that explicitly used interlayer edges, and it also indicates how the different layers in studies of multiple disconnected networks can be linked to each other. We have listed several types of layering, and it likely is also useful to explore other varieties.

**Supplementary Table 2:** We list a selection of diagnostics for, and other features of, multilayer networks; and we give example applications of these diagnostics and features for ecological multilayer networks (EMLNs).

| Diagnostic | Definition (words) | Example application(s) in EMLNs | Reference |
|---|---|---|---|
| Multiedge | A vector (whose length is equal to the number of layers) that consists of the set of edges that connect a given pair of nodes in different layers of a multilayer network. | Represent a set of ecological interactions between two species. | 109 |
| Multidegree | Total number of multiedges incident to a given node. | Quantify generality and/or vulnerability of a species across different interaction types, different places, etc. | 109 |
| Versatility | A generalisation of centrality measures that considers a node's importance across multiple layers. | Identify species that are important (i.e., versatile), because they bridge different interaction types, communities, and/or locations | 110 |
| Clustering coefficients | Measure extent to which nodes in a network tend to form transitive triples. | Evaluate the tendency of species to form interaction triples across sites or when there are multiple types of interactions. | 111 |
| Network community structure | Sets of nodes (in the same or different layers) that are connected to each other more densely than to other nodes in a network; network communities (i.e., modules) are usually identified as the output of an algorithm. | Identify modular structure across multiple times, spatial locations, and/or interaction types. | 50,112,113 |
| Stochastic block models | Generative statistical models to which one fits network data to find mesoscale features (such as modular structure, core–periphery structure, or others). | Identify sets of species across multiple times, spatial locations, and/or interaction types. | 114,115 |
| Node degree correlation across layers | Indicates relationship between intralayer degrees in different layers. (For example, are high-degree nodes in one layer typically also high-degree nodes in other layers?) | Are species generalists or specialists with respect to various interaction types or in various patches? | 116 |
| Reducibility | Find the layers needed to represent an ecological system in a way that is distinguishable from an aggregated network. | (1) Guide for temporal and/or spatial resolution at which data should be collected. (2) In a network in which each layer is a different community, indicates which of the communities contain unique species and/or interactions. (3) Informs researchers about aggregation of data. | 58 |
| Participation coefficient and overlapping degree | Structural role that a node plays in the distribution of edges across layers (in the spirit of the roles of species in interconnecting communities[4]). | Explore the functional role that species play in a community while considering several interaction types. | 117 |
| Node/interaction turnover | Mean proportion of node/edge appearance and disappearance in different layers. | Understand the time-dependence of network structure and the temporal turnover rates of species. | 8,118 |
| Measures of beta diversity | Similarity between a pair of network layers due to differences in species and/or interactions. | Investigate how networks vary in space or time, and connect such variation to ecological factors such as distance between patches or species' dispersal abilities. | 27 |

See the references for explicit mathematical definitions. Many other diagnostics for, and other features of, multilayer networks are discussed in Refs. [14,15].

**Supplementary Table 3:** Available software for analyzing multilayer networks

| Software | Interface | Uses | Link |
| --- | --- | --- | --- |
| MuxViz[119] | Graphical user interface (but may require some knowledge of R and Octave) | Visualization, interlayer correlations, network centrality, reducibility, modularity, node and edge statistics, motifs. | http://muxviz.net |
| Multilayer Networks Library | Python | Provides data structures for multilayer networks and basic methods for analysing them. | http://www.plexmath.eu/wp-content/uploads/2013/11/multilayer-networks-library_html_documentation/ |
| GenLouvain[77] | Matlab | Code for examining network community structure in multilayer networks with diagonal coupling. | http://netwiki.amath.unc.edu/GenLouvain/GenLouvain; in the SI, we include code (modified from the original) for the case in which layers have a bipartite structure. |
| LocalCommunityMatlab | | Local community-detection method used in ref.[112]. | https://github.com/LJeub/LocalCommunities |
| Infomap[113] | Stand-alone web interface | Community detection using flow-based ideas. | http://www.mapequation.org/code.html |
| Betalink package[27,120] | R | Quantify dissimilarity between ecological networks. | https://github.com/PoisotLab/betalink |
| timeordered package[13,121] | R | Analyse temporal networks. | https://cran.r-project.org/web/packages/timeordered/index.html |
| Multiplex package[122] | R | Algebraic procedures for the analysis of multiplex networks. Create and manipulate multivariate network data with different formats. | http://cran.r-project.org/web/packages/multiplex/index.html |

We specify websites to download the software. When relevant, we also include citations to associated references.

**Supplementary Table 4:** Community structure of the host–parasite temporal network for three examples of interlayer edge weights.

| Ecological Community | network type | Interlayer edge weights | $\overline{Q}_B$ | Number of Modules ($\overline{n}$) | Host adjustability, parasite adjustability |
|---|---|---|---|---|---|
| Layer 1 | monolayer | 0 | 0.46 | 6.47 | N/A |
| Layer 2 | monolayer | 0 | 0.43 | 5.63 | N/A |
| Layer 3 | monolayer | 0 | 0.40 | 6.22 | N/A |
| Layer 4 | monolayer | 0 | 0.49 | 4.16 | N/A |
| Layer 5 | monolayer | 0 | 0.35 | 4.49 | N/A |
| Layer 6 | monolayer | 0 | 0.56 | 5.88 | N/A |
| Whole network | multilayer | 1000 | 0.99 | 2.96 | 0 , 0 |
| Whole network | multilayer | Relative change in abundance | 0.55 | 6.30 | 0.474, 0.351 |

Each of the values of maximized modularity ($\overline{Q}_B$), the number of modules ($\overline{n}$), and the fraction of hosts and parasites that are assigned to more than one module (i.e., host and parasite 'adjustability', respectively) is a mean over 100 instantiations of the generalized Louvain algorithm[50,77]. We adjust the null model in the multilayer modularity objective function from Mucha et al.[50] to make it suitable for layers with a bipartite structure. In monolayer networks, it is not possible to compare modules across layers without resorting to an ad hoc approach, so we place a value of N/A in the table. We use the interlayer edge weight $\omega = 1000$ for our calculations near the extreme case $\omega = \infty$.

**Supplementary Table 5:** Comparison of the observed host–parasite network with weighted intralayer and interlayer edges to three null models.

| Shuffled | $\overline{Q}_B$ | Number of modules ($\overline{n}$) | Host adjustability | parasite adjustability |
|---|---|---|---|---|
| None (observed) | 0.55 | 6.3 | 0.474 | 0.351 |
| Intralayer edges | 0.41 ($P < 0.001$) | 4.9 ($P < 0.001$) | 0.372 ($P < 0.001$) | 0.448 ($P < 0.001$) |
| Interlayer edges: hosts | 0.21 ($P < 0.001$) | 38.6 ($P < 0.001$) | 1 ($P < 0.001$) | 0.928 ($P < 0.001$) |
| Interlayer edges: parasites | 0.21 ($P < 0.001$) | 94.6 ($P < 0.001$) | 1 ($P < 0.001$) | 0.928 ($P < 0.001$) |
| Permuted order of layers | 0.55 ($P \approx 0.52$) | 17.1 ($P < 0.001$) | 0.446 ($P < 0.001$) | 0.395 ($P < 0.001$) |

Each of the values of maximized modularity ($\overline{Q}_B$), the number of modules ($\overline{n}$), and the fraction of hosts and parasites that are assigned to more than one module (i.e., host and parasite adjustability, respectively) is a mean over 100 instantiations of the generalized Louvain algorithm[50,77]. We adjust the null model in the multilayer modularity objective function from Mucha et al.[50] to make it suitable for layers with a bipartite structure. We calculate the p-values $P$ (in parentheses) by comparing the mean of 100 instantiations of the observed network to 100 shuffled networks (see Supplementary Note 3 for details on how we shuffled networks).

**Supplementary Table 6:** Algorithms for examining network robustness to species extinctions.

| Removal scenario | Algorithm | Ecological reasoning/comments |
| --- | --- | --- |
| Monolayer: Remove plants and follow parasitoids' secondary extinctions. | **1.** Remove the plant with the lowest degree in the plant–parasitoid layer (in case of ties we order extinctions by species name). **2.** Remove all intralayer edges of the plant. **3.** Remove any parasitoids in the layer that remain disconnected (i.e., have degree 0). **4.** Record the proportion of surviving parasitoids in the layer as a function of the proportion of plants that have been removed. **5.** Repeat until all parasitoids are extinct in the layers. | **1.** Work only on the plant–parasitoid layer. **2.** We use proportions, because it allows us to compare the different scenarios. |
| Multilayer: Remove flower visitors and follow parasitoids' tertiary extinctions. | **1.** Remove the flower visitor with the lowest degree (in case of ties we order extinctions by species name). **2.** Remove all intralayer edges of the flower visitor. **3.** Remove any plants in the layer that remain disconnected (i.e., have degree 0). **4.** Remove any parasitoids that remain disconnected. **5.** Record the proportion of surviving plants and the proportion of surviving parasitoids (separately) as a function of the proportion of flower visitors that have been removed. **6.** Repeat until all parasitoids are extinct in the layers. | **1.** A removal of a plant from the flower-visitor layer entails immediate removal of a plant from the parasitoid layer. **2.** We use proportions, because it allows us to compare the different scenarios. |
| Multilayer: Remove flower visitors and follow parasitoids' tertiary extinctions; we also include uniformly random plant removal | **1.** Remove the flower visitor with the lowest degree (in case of ties we order extinctions by species name). **2.** Remove all intralayer edges of the flower visitor. **3.** Remove any plants in the layer that remain disconnected (i.e., have degree 0). **4.** Select a plant from the surviving plants uniformly at random and remove it with a probability of $p_{\text{rand}}$. **5.** Remove any parasitoids that remain disconnected. **6.** Record the proportion of surviving plants and the proportion of surviving parasitoids (separately) as a function of the proportion of flower visitors that have been removed. **7.** Repeat until all parasitoids are extinct in the layers. | **1.** Same as in the previous scenario but incorporates a uniformly random plant extinction with probability $p_{\text{rand}}$ (we consider examples with $p_{\text{rand}} = 0.3$ and $p_{\text{rand}} = 0.8$) to acknowledge the fact that extinctions can happen not only from removal of flower visitors but also from other processes that operate directly on plants. **2.** Because of the random plant extinctions, we average results over 100 instantiations. |

# 1 Supplementary Note 1: Multilayer Modularity Maximization

We detect communities using a multilayer modularity quality function $Q_{\text{multislice}}$[50]. Consider a (weighted or unweighted) unipartite multilayer network whose only interlayer edges are diagonal (i.e., interlayer edges only connect nodes to their counterparts across layers). With the standard (Newman–Girvan) null model, multilayer modularity is[50]

$$Q_{\text{multislice}} = \frac{1}{2\mu} \sum_{ijsr} \left[ \left( A_{ijs} - \gamma_s \frac{k_{is} k_{js}}{2m_s} \right) \delta_{sr} + \delta_{ij} \omega_{jsr} \right] \delta(g_{is}, g_{jr}), \qquad (1)$$

where $A_{ijs}$ is the weight of the intralayer edge between nodes $i$ and $j$ on layer $s$ (for an unweighted network, the weight is exactly 1 if there is an edge), the tensor element $\omega_{jsr}$ gives the weight of the interlayer edge between node $j$ on layer $r$ and node $j$ on layer $s$, the resolution-parameter value on layer $s$ is $\gamma_s$, the quantity $k_{is}$ is the intralayer strength (i.e., weighted degree) of node $i$ on layer $s$ (and $k_{js}$ is defined analogously), $2m_s$ is the total edge weight in layer $s$, the set $g_{is}$ is the community that includes node-layer $(i, s)$ (and $g_{js}$ is defined analogously), we denote the Kronecker delta between indices $x$ and $y$ by $\delta_{xy}$ (which is equal to 1 for $x = y$ and equal to 0 for $x \neq y$), and $2\mu = \sum_{ijs} A_{ijs}$. In an ordinal multilayer network (e.g., for the usual multilayer representation of a temporal network), $\omega_{jsr}$ can be nonzero only when $s$ and $r$ are consecutive layers.

For our calculations, we need to adjust the null-model contribution $P_{ijs} = \gamma_s \frac{k_{is} k_{js}}{2m_s}$, which gives the expected number of interactions between nodes $i$ and $j$ in layer $s$, to consider the bipartite structure of a host–parasite network. That is, hosts are only allowed to be adjacent to parasites, and parasites are only allowed to be adjacent to hosts. (See refs.[51,52,123] and references therein for discussions of null models for multilayer modularity maximization.) To do this, we adapt Eq. (15) from Barber *et. al*[124] and write a bipartite multilayer modularity as

$$Q_B = \frac{1}{2\mu} \sum_{ijsr} \left[ \left( A_{ijs} - \gamma_s \frac{k_{is} d_{js}}{m_s} \right) \delta_{sr} + \delta_{ij} \omega_{jsr} \right] \delta(g_{is}, g_{jr}), \qquad (2)$$

where $k_{is}$ is the strength of host $i$ on layer $s$ and $d_{js}$ is the strength of parasite $j$ on layer $s$. We modify the GENLOUVAIN code[77] from http://netwiki.amath.unc.edu/GenLouvain/GenLouvain (see Supplementary Table 2 and the MATLAB code in the Supplementary Information) by changing the null model $P_{ijs}$. We use the default resolution-parameter value of $\gamma = 1$[50,51].

We maximize the quality function (2) using a generalized Louvain-like locally greedy algorithm[77] (see[51] for details on the generalized Louvain algorithm and[125] for the original Louvain algorithm), which also determines module affiliation[50,51]. As with all other algorithms for maximizing modularity in monolayer networks[48,49,126], the Louvain algorithm is a heuristic, and it yields a network partition that corresponds to a local maximum of modularity (and, importantly, we note that the modularity landscape has a very large number of "good" local maxima)[127]. It is also a stochastic algorithm, so different results can (and generally do) yield different module assignments and somewhat different modularity values. We thus average the values of the maximized modularity $Q_B$ and the number $n$ of modules across 100 instantiations of the algorithm and present results for the mean modularity $\overline{Q}_B$ and mean number $\overline{n}$ of modules.

The generalized Louvain algorithm[77] receives as input a modularity matrix, which is calculated according to the supra-adjacency matrix of a network. (See Eq. (2), and Supplementary Fig. 1b for an example of a supra-adjacency matrix.) Given any supra-adjacency matrix that represents a diagonally-coupled multilayer network, we have a viable input for optimizing multilayer modularity of that network. The output of one instantiation of the generalized Louvain algorithm[77] algorithm is a value for $Q_B$ and an assignment of each state node to a module.

## 2 Supplementary Note 2: Analysis of a Synthetic Network with Two Interaction Types

The network in Fig. 2 has a 'planted' modular structure of 3 modules in each layer. We use a binary (i.e., unweighted) form of the network, in which intralayer edges have a value of 1 if they exist and 0 if they do not[10]. For simplicity, we assume that all interlayer edges have the same weight. To present the results of the optimization procedure, we report mean values $(\overline{Q}_B, \overline{n})$, which we obtain by averaging over 100 instantiations of the modularity-maximization algorithm, for multilayer modularity $Q_B$ and the number $n$ of modules.

## 3 Supplementary Note 3: Analysis of a Temporal Network

### 3.1 Interlayer Edge Weights on a Comparable Scale as Intralayer Edges.

We examine a temporal network that encodes the infection of 22 small mammalian host species by 56 ectoparasite species (mites and fleas) during six consecutive summers in Siberia (1982–1987), giving 6 different layers. (See Refs.[54,55] for more details on data.) Not all species occur in all years. The data set also includes the number of host specimens collected in the field and the number of parasites recovered from each host specimen. We use these individual-level estimates of host and parasite abundances determine, respectively, the weights of intralayer and interlayer edges. The weight of an intralayer edge between host $i$ and parasite $j$ at time $t$ is $A_{ijt} = m_{ij}^t = p_{ij}^t/a_i^t$, where $A_{ijt}$ denotes an intralayer edge weight in layer $t$, the quantity $p_{ij}^t$ is the number of parasite specimens of species $j$ recovered from all of the specimens of host $i$ at time $t$, and $a_i^t$ is the number of specimens of host $i$ at time $t$. The interlayer edge that connects host species $i$ at time $t$ to its counterpart at time $t+1$ has a weight $\omega_{it,t+1} = n_i^{t \to t+1} = a_i^{t+1}/a_i^t$ given by the relative change in host abundance between the time points. An interlayer edge from node $i$ in layer $r$ to node $i$ in layer $s$ has weight $\omega_{irs}$, where we have included a comma in the subscript between $r = t$ and $s = t+1$ for clarity and where we take $\omega_{irs} = \omega = $ constant in the synthetic network example. Analogously, the interlayer edge that connects parasite species $j$ at time $t$ to its counterpart at time $t+1$ has a weight $\omega_{jt,t+1} = n_j^{t \to t+1} = p_j^{t+1}/p_j^t$ given by the relative change in the total abundance of the parasite (across all hosts). We do not include an interlayer edge between corresponding species in layers $t$ and $t+1$ when (i) a species occurs in layer $t$ but not in layer $t+1$ or (ii) when a species does not occur in layer $t$ but occurs in layer $t+1$.

The above formulation of edge weights is ecologically meaningful, as the weights reflect temporal dynamics in species abundances. It also allows one to model how changes in species abundances across time affect host–parasite interactions in each time point. For example, the future infection of a host $i$ with parasite $j$ depends on the interlayer edges and abundance values as follows:

$$m_{ij}^{t,t+1} = \frac{p_{ij}^{t+1}}{a_i^{t+1}} = \frac{p_{ij}^{t+1}}{n_i^{t \to t+1} a_i^t} = \frac{n_j^{t \to t+1} p_j^t}{n_i^{t \to t+1} a_i^t}. \qquad (3)$$

Although we do not use the information in Eq. (3), it may be helpful for researchers who are interested in various questions. For example, one can ask how does reducing the abundance of a given host (e.g., by control programs) will affect future interactions of the host with parasites, or, more generally, network architecture. The formulation of edge weights gives nonzero weights to non-diagonal interlayer edges, which was done previously (and very effectively) for temporal networks of disease propagation in Valdano *et al.*[128].

It will also be instructive, for the sake of comparison with the two extreme scenarios $\omega = 0$ and $\omega = \infty$ (see Section 3.2), to indicate the numerical values of the weights of interlayer edges, which vary between 0.02 and 10.33. We show the distribution of the interlayer edge values in Supplementary Fig. 4.

For simplicity, in all of our calculations of the host–parasite temporal network, we follow previous studies and assume that the intralayer and interlayer[50,51,56] edges are undirected. It will be interesting to relax this assumption in future studies of multilayer temporal networks with interlayer connectivity.

### 3.1.1 Testing Hypotheses About the Modular Structure

We now compare the modular structure in the observed network to that obtained from three different null models[19,62]. For each null model, we test the hypothesis that the most modular partition of the observed temporal network is more modular than the most modular partition of shuffled networks, and we calculate the significance of the modularity $Q_B^{OBS}$ of the observed network's most modular partition as the proportion of the 100 shuffled networks that have a lower maximized modularity than that of the observed network[56]. We also calculate the number of modules and the 'adjustability' of hosts and parasites as the proportion of nodes of a given type that change modules at least once. For comparative applications of adjustability, we take into account the total number of temporal layers. We compare our results with shuffled networks to the values that we obtain for the observed network using one-sample $t$-tests[19].

In the test with the first null model, we shuffle the intralayer edges in each layer to test the hypothesis that modular structure over time depends on temporally-local interaction patterns. In each layer, we first shuffle the total number of parasites that are on a given host. We use the R0_BOTH algorithm from the VEGAN package[129] in R. The algorithm shuffles the values and position of the matrix entries within each row, while preserving the total sum of the matrix. It therefore conserves both the number of parasite species that infect a given host and the mean abundance of those species. We then divide each matrix entry of each shuffled matrix by the host abundance in that particular layer (i.e., year) to transform the abundance to prevalence values. We find statistically significant support for the hypothesis (see Supplementary Table 5). Additionally, the observed network differs in a statistically significant way from the shuffled networks with respect to both number of modules and host and parasite adjustability (see Supplementary Table 5).

In the second test, we shuffle the interlayer edges between each pair of consecutive layers to test the hypothesis that the modular structure depends on the identity of hosts and parasites that change in abundance. To construct this null model, which is equivalent to the 'nodal' null model from Bassett et al.[19], we separately permute the orders of node labels of hosts and parasites. We permute the node labels separately in each layer so that node identity is not preserved. In the bipartite structure of a given layer, this is akin to permuting the order of rows (for hosts) or columns (for parasites) in each layer. We find that (i) the most modular partition of the observed network is more modular than the most modular partition of the shuffled networks (see Supplementary Table 5) and (ii) their temporal structures differ. Specifically, the most modular partition of the observed network has approximately 6 times fewer modules than when shuffling the order of hosts and approximately 15 times fewer modules than when shuffling the order of parasites. Additionally, in the shuffled networks, almost all hosts and parasites switch modules at least once (see Supplementary Table 5).

In the third test, we permute the order of the layers to test the hypothesis that the order of time affects the network's modular structure. To construct this null model, which is equivalent to the 'temporal' null model of Bassett et al.[19], we draw a uniformly random permutation of the sequence $\{1,\ldots,6\}$ (e.g., (4,3,2,5,1,6)) to give a new ordering of the layers. We thereby destroy the temporal identity of the layers. Note that the interlayer edges are not shuffled. We find that the most modular partition of the observed network is not more modular than the most modular partition of the randomized networks (see Supplementary Table 5), but that the observed and randomized networks have different temporal structures. The most modular partition of the observed network has about 2.5 times fewer modules than those of the randomized networks, and it also has different proportions of hosts and parasites that switch modules (see Supplementary Table 5).

## 3.2 Interlayer Edge Weights of 0 or Infinity

We compare the modular structure of the network with weighted intralayer and interlayer edges to the modular structure of two versions of the same network but in which interlayer edge values are set to either 0 or are very large. We perform these computations in the same manner as we did for the interlayer-weighted network that we discussed in detail in Section 3.1. We again use the generalized Louvain algorithm from Jutla et al.[77] and Mucha et al.[50]. As a proxy for the scenario with infinitely large interlayer edge weights using numerical calculations, we set $\omega = 1000$. (Technically, we are thus approximate in our exploration of this extreme situation.) To explore the scenario with 0 interlayer edge weights, we examine 6 independent modularity matrices (corresponding to the 6 layers). We also provide the code for these computations.

**Very large interlayer edge weights.** When the weight of all interlayer edges is very large, processes that govern interactions across contiguous layers (e.g., species persistence, survival, or changes in abundance) have a much larger effect on community structure than those that govern interactions within layers. Because the effect of species connectivity across time is much stronger than that of the intralayer processes, we expect species to remain in the same module over time. In the limit of infinite-weight interlayer edges for all species, all state nodes that correspond to a given species must always be assigned to the same module. In our computation with interlayer edge weights of $\omega = 1000$, we obtain $\overline{Q}_B \approx 1$, and the maximum-modularity partition of the network has $\overline{n} = 3$ modules. By construction, no species changes its module affiliation across time (see Supplementary Table 4).

**Interlayer edge weights of** 0. At the other extreme is a multilayer network in which all of the interlayer edges have a weight of 0, which is akin to maximizing modularity separately in each of the layers. In this case, the modules that one finds in different layers are independent of each other, as they reflect only the internal structure of each layer[50,51]. This is the same situation as calculating modularity for two disconnected networks in the toy example (see Fig. 2b). From an ecological perspective, one cannot understand the effect of interlayer phenomena (e.g., temporal changes in species abundance or in network architecture) on modular structure in a direct manner if one has exclusively 0 interlayer edge weights. Although some quantification of module evolution may be possible[53], such a quantification does not simultaneously encompass processes that occur within and between layers. For example, if a species interacts strongly with a given set of species in layer $t$ but with another set of species in layer $t+1$, one can argue that this is a change in the modular structure over time. However, even though it is possible, it need not be true, because the modules have been calculated independently for two distinct networks. Moreover, if a community splits in time, including nonzero interlayer edges provides a natural way to track which nodes leave to form a new community and which remain part of the original community.

When all interlayer edge weights are 0, we find that both $\overline{Q_B}$ and $\overline{n}$ vary among the 6 layers. (See Supplementary Table 4 for the exact values.) It is not possible to calculate changes in module composition (including host and parasite adjustability) in an unambiguous way, because modules are defined separately in a layer. One can, of course, attempt ad hoc analyses of modular structure changes in time even when interlayer edge weights are 0.

### 3.3 Calculation of Normalized Mutual Information

To support the claim that modularity maximization in a multilayer network provides additional information about the assignment of species to modules compared to independent computations in all individual layers, we quantify the amount of novel information from the addition of interlayer edges for the assignment of species to modules. To do this, we use a measure of normalized mutual information (NMI) based on information theory. This quantity has been used previously to evaluate methods that assign nodes to modules[130,131]. For each of the six layers, the value of NMI is 1 if the composition of modules is identical in the monolayer and the multilayer networks, and it is 0 if they are completely uncorrelated. That is, a value of 1 means that one has complete information on the assignment of species to modules in the temporal network from independent computations with the corresponding layer. In that case, the interlayer edges have not given any additional information. In contrast, a value of 0 means that observing a sin-gle layer at a time provides no information whatsoever on the assignment of species to modules in the multilayer representation of the temporal network. Naturally, the only way to know if one gets the same answer is to perform the multilayer calculation in the first place. We find that across layers the mean NMI ($\pm$ standard deviation) is $0.35 \pm 0.12$. (Thus, the range of values is $0.19--0.51$, and we recall that there are $M = 6$ layers.) Hence, the calculation of modularity of each layer by itself contains, on average, only about 35% of the information on species assignment to modules in the multilayer network with the interlayer edges. Combined with the fact that we can use the interlayer edges at time $t-1$ to calculate the intralayer edges at time $t$ (see Eq. 3), this gives compelling evidence that including interlayer edges provides new information on the temporal organization of the network.

We also calculate the amount of information that is contained in the assignment of species to modules in an aggregated network compared to module assignments in any individual layer. We use two versions of an aggregated multilayer network: (i) we sum the intralayer edge weights (i.e., interaction values) across the 6 layers; and (ii) we take the mean of the edge weights across the 6 layers. We find on average that the correlation in the information is about $0.521 \pm 0.029$ and $0.517 \pm 0.032$ for the two types of aggregated networks, respectively. Consequently, the affiliation of species to modules in either of these aggregated networks provides only about 52% of the information on their affiliation to modules in the multilayer network (which includes interlayer edges).

### 3.4 Reducibility of the Host–Parasite Network

To examine if we need all 6 layers in the host–parasite network to accurately describe structure in the host–parasite network, we calculate *reducibility* of the network[58]. Briefly, reducibility is a measure of the structural redundancy of different layers with respect to an aggregated network $G = G^{[1]} + G^{[2]} + \cdots + G^{[M]}$, where $\{G^{[1]}, G^{[2]}, \ldots, G^{[M]}\}$ is the set of adjacency matrices from the $M$ individual layers of a multilayer network. Intuitively, if all of the layers are the same, then there is complete redundancy, and a single layer includes all of the information about intralayer structure. However, if each layer includes different intralayer edges, then we need to retain all layers to fully describe structure in a multilayer network, and aggregation leads to loss of information.

From our computation, we see for our temporal network that we need all 6 layers to accurately describe its structure (see Supplementary Fig. 6). More generally, temporal interaction dynamics are accompanied by a temporal ordering of interactions, which are also often bursty, so aggregating temporal layers into a time-independent network (especially if done naively, as is common) removes all information about the ordering of interactions and can lead to qualitatively incorrect conclusions[57,132].

# 4 Supplementary Note 4: Further Details on Robustness Analysis

We use empirical data from Pocock et al.[38], who assembled a network composed of 10 layers, which are connected to each other mostly via plants. Layers in the original network include interactions of plants with granivorous birds, rodents, butterflies and other flower visitors, aphids, granivorous insects, and leaf-miner parasitoids. For simplicity, we select a subset of the 10 layers and form a diagonally-coupled multilayer network with two layers. Our network describes two interconnected mutualistic interactions, although the mutualism in the leaf-miner parasitoid layer is indirect. The network includes 189 flower visitors, 60 leaf-miner parasitoids, and 17 plants. For simplicity, we use a binary version of the original data.

Pocock et al.[38] studied the effect of plant removal on the simultaneous co-extinction of the species that are interconnected to them. However, they did not study how disturbance percolates through plants, and they thus did not explore how the multilayer structure of their network affects cascading perturbations. In the context of our example, extinction of pollinators entails co-extinction of plants, and extinction of plants entails co-extinction of leaf-miner parasitoids. Because an extinction of plants in one layer automatically implies their extinction in the second layer, we assume that the weights of interlayer edges are infinity. We modify an established approach for examining co-extinctions[38,133] so that we can use it with multilayer networks (see Supplementary Table 6).

We explore another option to compare the robustness of a plant–parasitoid monolayer network (which is essentially one layer of the multilayer network) to that of a flower visitor–plant–parasitoid multilayer network. Unlike in the main text, we remove (in the same order) the same set of nodes (flower visitors) in the monolayer network and in the multilayer network, and we track the total proportion of species that remain in each of these networks. Specifically, in the monolayer network, we remove flower visitors in order of increasing degree and then compute the proportion of plants that remain. In the multilayer network, we remove the same flower visitors in the same order and then track the proportions of both plants and parasitoids that remain. In the multilayer network, we find that the total proportion of surviving species is larger and that the pace of extinction is slower (see Supplementary Fig. 5).

Like standard simple approaches for studying network robustness[63], our calculations have the flavour of a percolation process. See Nagler et al.[134] for an example of a percolation-based robustness study of an EMLN. Multilayer networks allow one to examine a much richer variety of percolation processes than in monolayer networks[14,15,116], and that in turn allows increasingly nuanced analyses of species robustness in ecological networks in a wealth of scenarios.